\documentclass[11pt,a4paper]{article}
\usepackage[margin=2.5cm]{geometry}
\usepackage{amsmath,amssymb,graphicx,booktabs,xcolor,mathrsfs}
\usepackage[colorlinks=true,linkcolor=blue!60!black,citecolor=blue!60!black,urlcolor=blue!60!black]{hyperref}
\usepackage[font=small,labelfont=bf]{caption}

\newcommand{\doilink}[1]{\href{https://doi.org/#1}{\texttt{doi:#1}}}
\newcommand{\arxivlink}[1]{\href{https://arxiv.org/abs/#1}{\texttt{arXiv:#1}}}
\newcommand{\inspirelink}[1]{\href{https://inspirehep.net/literature?q=j+#1}{[INSPIRE]}}

\newcommand{\GR}{\mathscr{G}_R^Z}

\title{Parity restoration in a sterile-neutrino scenario: bounds from atomic parity violation to the LHC}
\author{Iman Motie\textsuperscript{a}, {She-Sheng Xue}\textsuperscript{b--e}, Alain Blanchard\textsuperscript{a}\\[4pt]
\small {a}) Universit\'e de Toulouse, UPS-OMP, IRAP, F-31400 Toulouse, France\\
\small b)  ICRANet, Piazza della Repubblica 10, I-65122 Pescara, Italy\\
\small c) Physics Department, Sapienza University of Rome, Rome, Italy\\
\small d) INFN, Sezione di Perugia, Perugia, Italy\\
\small e) ICTP-AP, University of Chinese Academy of Sciences, Beijing, China}
\date{}

\begin{document}
\maketitle

\begin{abstract}
In a four-fermion-interaction scenario beyond the Standard Model (SM), in which
right-handed (sterile) neutrinos take part in TeV-scale composite dynamics, the
$Z$ boson acquires an energy-dependent right-handed coupling $\GR(q^2)$ that is
expected to vanish at low energies and to restore parity symmetry at high
energies. We derive the contribution of $\GR$ to neutral-current parity-violating
observables. At first order in $\GR$, every low-energy parity-violating coupling
is obtained from its SM value by the replacement
$\sin^2\theta_W\to\sin^2\theta_W(1-\GR)$; in particular the nuclear weak charge
is shifted by $\Delta Q_W=4\sin^2\theta_W\,Z\,\GR$, while the $Z$-pole effective
mixing angle is shifted by $-\tfrac12(1-2\sin^2\theta_W)\GR$. Using atomic parity
violation in cesium, parity-violating electron scattering and $Z$-pole
determinations of $\sin^2\theta_{\rm eff}$, we extract $\GR$ at each momentum
transfer. All data are consistent with $\GR=0$; at the $Z$ pole we find
$\GR(m_Z)<1.1\times10^{-3}$ (95\% CL). We then use the CMS measurement of the
Drell--Yan forward--backward asymmetry at dilepton masses above 170~GeV to
constrain the coupling directly at TeV momentum transfer, finding $\GR<0.04$
(95\% CL) for $0.2<\sqrt{q^2}<2$~TeV. Combining both, the scale of parity
restoration in a running $\GR=q^{2n}/(q^{2n}+\Lambda^{2n})$ must satisfy
$\Lambda>3.8$, $2.1$ and $1.7$~TeV for $n=1,2,3$. Future MOLLER, P2 and HL-LHC
data will improve these bounds.
\end{abstract}

\section{Introduction}\label{sec1}

Symmetry principles have long played a crucial role in understanding fundamental
interactions. Among them, parity, the inversion of spatial coordinates, occupies
a special place: it was historically assumed to be exact, and its violation in
weak interactions revealed deep structural features of the Standard Model (SM).
The possibility that parity is not conserved in weak interactions was raised by
Lee and Yang~\cite{LeeYang}, and was confirmed in 1957 by Wu and collaborators in
the $\beta$ decay of polarized $^{60}$Co~\cite{Wu}. These discoveries established
that charged-current weak interactions maximally violate parity, motivating
decades of precision tests in nuclear and atomic systems.

In 1973 neutral weak currents, mediated by the $Z^0$ boson, were discovered at
CERN~\cite{Haidt}. Parity violation in the neutral-current channel was first
observed by Prescott \emph{et al.}\ at SLAC in deep-inelastic scattering of
polarized electrons on deuterium~\cite{Prescott}, opening the way to precision
determinations of the weak mixing angle and to searches for new physics in
neutral-current interactions.

In atoms, the parity-violating interaction between atomic electrons and the
nucleus is characterized by the nuclear weak charge $Q_W$, the coupling of the
nucleus to the $Z^0$ boson. Parity-violating effects grow roughly as $Z^3$, which
makes heavy atoms particularly sensitive to small deviations from the SM. Cesium
offers a good compromise between a large atomic number and a relatively simple
atomic structure, which allows atomic-structure calculations at the sub-percent
level. The $6S$--$7S$ transition in cesium~\cite{Wood} provides the most precise
determination of $Q_W$ and one of the most precise low-energy determinations of
the weak mixing angle. Reviews of atomic parity violation (APV) can be found in
Refs.~\cite{ErlerSu,Bouchiat81,Bouchiat97,RamseyMusolf,Rosner}.

In this work we study how the right-handed neutral-current coupling $\GR$,
predicted, together with three sterile neutrinos and parity-symmetry restoration,
by the four-fermion-interaction scenario beyond the SM of
Refs.~\cite{Xue2016,Xue2017,Xue2022}, modifies neutral-current parity-violating
observables. Our aims are (i) to derive the shift of the nuclear weak charge and
of the effective weak mixing angle at first order in $\GR$; (ii) to extract $\GR$
from measurements at momentum transfers from a few MeV (APV) to the $Z$ pole;
(iii) to constrain it directly at TeV momentum transfer with the high-mass
Drell--Yan forward--backward asymmetry measured by CMS; and (iv) to translate these into constraints on the energy dependence of $\GR(q^2)$
and on the scale $\Lambda$ at which parity symmetry is expected to be restored.

The paper is organized as follows. Section~\ref{sec2} reviews the neutral current
and APV in the SM. Section~\ref{sec3} introduces the right-handed coupling and
derives its contribution to the weak charge. Section~\ref{sec4} treats the
effective weak mixing angle and the left--right asymmetry at the $Z$ pole and
presents the data-driven extraction of $\GR(q^2)$. Section~\ref{sec5} derives
TeV-scale constraints from the high-mass Drell--Yan forward--backward asymmetry,
Section~\ref{sec6} discusses future probes, and Section~\ref{secsumm}
summarizes.

\section{The neutral current and atomic parity violation}\label{sec2}

The coupling of the $Z$ boson to a SM fermion $f$ with third isospin component
$T_3^f$ (of its left-handed component) and electric charge $Q_f$ is
\begin{equation}
\mathcal{L}_{\rm NC}= \frac{g}{c_W}\,Z_\mu \sum_f \bar f\gamma^\mu
\left(g_V^f-g_A^f\gamma_5\right) f,
\qquad g_V^f=\tfrac12T_3^f-Q_f\,s_W^2,\quad g_A^f=\tfrac12T_3^f,
\label{eq:LNC}
\end{equation}
with $s_W\equiv\sin\theta_W$, $c_W\equiv\cos\theta_W$ and $g\,s_W=g'c_W=e$. The
quantum numbers are listed in Table~\ref{tab1}. Explicitly, for the first
generation the neutral current reads
\begin{align}
J_Z^\mu ={}& \tfrac14\,\bar e\gamma^\mu\gamma_5 e+\big(-\tfrac14+s_W^2\big)\bar e\gamma^\mu e
-\tfrac14\,\bar u\gamma^\mu\gamma_5 u+\big(\tfrac14-\tfrac23 s_W^2\big)\bar u\gamma^\mu u \nonumber\\
&+\tfrac14\,\bar d\gamma^\mu\gamma_5 d+\big(-\tfrac14+\tfrac13 s_W^2\big)\bar d\gamma^\mu d .
\label{eq:JZ}
\end{align}

\begin{table}[t]
\centering
\begin{tabular}{lcccc}
\toprule
$f$ & $T_3$ & $Q$ & $Y_R$ & $Y_L$\\
\midrule
$\nu_e,\nu_\mu,\nu_\tau$ & $+\tfrac12$ & $0$ & $0$ & $-\tfrac12$\\
$e,\mu,\tau$ & $-\tfrac12$ & $-1$ & $-1$ & $-\tfrac12$\\
$u,c,t$ & $+\tfrac12$ & $+\tfrac23$ & $+\tfrac23$ & $+\tfrac16$\\
$d,s,b$ & $-\tfrac12$ & $-\tfrac13$ & $-\tfrac13$ & $+\tfrac16$\\
\bottomrule
\end{tabular}
\caption{Neutral-current quantum numbers of the SM fermions ($Q=T_3+Y$).}
\label{tab1}
\end{table}

At momentum transfers $|q^2|\ll m_Z^2$, exchange of a $Z$ boson between an
electron and a quark gives the four-fermion interaction
\begin{equation}
\mathcal{L}_{\rm eff}=\frac{G_F}{\sqrt2}\sum_{q=u,d}\Big[
C_{1q}\,\bar e\gamma^\mu\gamma_5 e\;\bar q\gamma_\mu q
+C_{2q}\,\bar e\gamma^\mu e\;\bar q\gamma_\mu\gamma_5 q\Big],
\label{eq:Leff}
\end{equation}
where we used $g^2/(c_W^2m_Z^2)=e^2/(s_W^2c_W^2m_Z^2)=4\sqrt2\,G_F$ and
\begin{equation}
C_{1q}=8\,g_A^e g_V^q,\qquad C_{2q}=8\,g_V^e g_A^q ,
\label{eq:C1C2}
\end{equation}
i.e.\ at tree level $C_{1u}=-\tfrac12+\tfrac43s_W^2$,
$C_{1d}=\tfrac12-\tfrac23s_W^2$ and $C_{2u}=-C_{2d}=-\tfrac12+2s_W^2$. In
nuclei the quark vector currents add coherently, so that APV is dominated by the
$C_{1q}$ terms (axial electron current times vector quark current), whereas the
$C_{2q}$ terms are spin dependent and suppressed. The resulting effective
Hamiltonian for an atomic electron is~\cite{BouchiatFayet}
\begin{equation}
H_{\rm eff}=\frac{G_F}{2\sqrt2}\,Q_W(Z,N)\,e^\dagger(\vec r)\gamma_5 e(\vec r)\,\rho(\vec r),
\end{equation}
where $\rho$ is the nuclear density and the weak charge of a nucleus with $Z$
protons and $N$ neutrons is
\begin{equation}
Q_W(Z,N)=-2\big[(2Z+N)C_{1u}+(Z+2N)C_{1d}\big]=Z(1-4s_W^2)-N .
\label{eq:QW}
\end{equation}

A generic new parity-violating contact interaction~\cite{Arcadi}
\begin{equation}
\mathcal{L}^{\rm NP}_{\rm eff}=\frac{1}{\Lambda^2}\,\bar e\gamma^\mu\gamma_5e
\big(f_{Vu}\,\bar u\gamma_\mu u+f_{Vd}\,\bar d\gamma_\mu d\big)
\label{eq:LNP}
\end{equation}
shifts $C_{1q}\to C_{1q}+\sqrt2 f_{Vq}/(G_F\Lambda^2)$, and therefore
\begin{equation}
\Delta Q_W\equiv Q_W^{\rm exp}-Q_W^{\rm SM}
=-\frac{2\sqrt2}{G_F\Lambda^2}\big[(2Z+N)f_{Vu}+(Z+2N)f_{Vd}\big]
=-\frac{2\sqrt2}{G_F}\,\frac{f^{\rm eff}_{Vq}}{\Lambda^2}\,3(Z+N),
\label{eq:dQW}
\end{equation}
with $f^{\rm eff}_{Vq}\equiv[f_{Vu}(2Z+N)+f_{Vd}(Z+2N)]/[3(Z+N)]$.

For $^{133}_{\;\,55}$Cs the SM predicts $Q_W^{\rm SM}=-73.26(1)$~\cite{PDG2024}.
The extracted experimental value depends on the atomic-structure calculation
used to interpret the measured amplitude: Ref.~\cite{Dzuba} obtains
$Q_W^{\rm exp}=-72.58(29)_{\rm exp}(32)_{\rm th}$, while Ref.~\cite{Porsev} obtains
$-73.16(29)_{\rm exp}(20)_{\rm th}$. With the former,
$\Delta Q_W=0.68\pm0.43$, i.e.\ $-0.16<\Delta Q_W<1.52$ at 95\% CL, and
Eq.~\eqref{eq:dQW} with $G_F=1.1663788\times10^{-5}$~GeV$^{-2}$ gives
\begin{equation}
-1.6\times10^{-8}\ {\rm GeV}^{-2}<\frac{f^{\rm eff}_{Vq}}{\Lambda^2}<1.7\times10^{-9}\ {\rm GeV}^{-2}
\qquad(95\%\ {\rm CL}).
\label{cons}
\end{equation}
With the convention of Eq.~\eqref{eq:LNP}, a positive $\Delta Q_W$ corresponds to a negative $f^{\rm eff}_{Vq}$.

\section{Right-handed neutral-current coupling}\label{sec3}

\subsection{Motivation}
In the scenario of Refs.~\cite{Xue2016,Xue2017}, the SM four-fermion
interactions, which involve three right-handed (sterile) neutrinos, become strong
at TeV scales and bind SM fermions into massive composite particles. The SM gauge
bosons and the composite particles restore parity-preserving gauge symmetries at
the ultraviolet fixed point of the theory~\cite{Xue2013,Xue2016b}, while the
parity-violating SM is recovered at the infrared fixed point,
$v\simeq246$~GeV. As a consequence the $W$ and $Z$ bosons acquire non-trivial,
energy-dependent right-handed gauge couplings $\mathscr{G}_R^{W}(q^2)$ and
$\GR(q^2)$, obtained from the corresponding self-energies~\cite{Xue2022}:
$\mathscr{G}_R^{W,Z}\to0$ at low energies, recovering the SM, and
$\mathscr{G}_R^{W,Z}\to1$ at high energies, where parity symmetry is restored.

In this scenario the SM is extended by three right-handed (sterile) neutrinos
$\nu_R^{e,\mu,\tau}$, so that every SM fermion has both chiralities and the
four-fermion operators of Nambu--Jona-Lasinio type,
$G_{\rm cut}\,(\bar\psi_L^{i}\psi_{Rj})(\bar\psi_R^{j}\psi_{Li})$, can be written
for leptons as well as quarks~\cite{Xue2016,Xue2017}. When such an operator is
inserted into the one-loop $Z$ vertex, the gauge boson couples to the
left-handed current of the loop fermions, but the operator turns that pair into
an external right-handed pair. The result is an induced right-handed vertex,
$\GR(q^2)\,T_3^f\,\bar f\gamma^\mu P_R f$, whose strength is fixed by the
four-fermion coupling and by the fermion self-energy $\Sigma(q^2)$ generated by
the same interaction~\cite{Xue2022}. Without $\nu_R$ the lepton sector has no
such operators and no right-handed $Z$ coupling. Its presence makes $\GR$
universal for quarks and leptons, which is what we assume in
Secs.~\ref{sec4}--\ref{sec5}. Below the electroweak scale the four-fermion
coupling flows to its infrared fixed point and $\GR\to0$. Near the
ultraviolet fixed point, where SM fermions and $\nu_R$ form composite Dirac
states with vector-like gauge couplings, $\GR\to1$.

This scenario was originally invoked~\cite{Xue2022} in connection with the
$W$-mass measurement of CDF~\cite{CDF}, which deviated from the SM by $7\sigma$.
Subsequent measurements by ATLAS~\cite{ATLASW} and CMS~\cite{CMSW} are consistent
with the SM and in tension with the CDF value, so that the $W$ mass no longer
calls for a large right-handed coupling at the electroweak scale. Constraints on
the charged-current coupling $\mathscr{G}_R^W$ and on the sterile neutrinos come
from vector-boson fusion and heavy-neutral-lepton searches at the
LHC~\cite{CMSVBF,CMSHNL1,CMSHNL2}, from neutrinoless double-beta
decay~\cite{Pacioselli}, from the ratio of top- and bottom-quark
masses~\cite{Xue2016b}, from CMB birefringence~\cite{Mahmoudi} and from the muon
anomalous magnetic moment~\cite{Motie}, indicating $\mathscr{G}_R^W\lesssim10^{-4}$
at low energies. The neutral-current coupling $\GR$, the subject of this paper, is
probed independently by parity-violating neutral-current observables.

\subsection{Effective couplings}
After electroweak symmetry breaking, the right-handed coupling modifies the $Z$
vertex of Eq.~\eqref{eq:LNC} as
\begin{equation}
\mathcal{L}_{\rm NC}=\frac{g}{c_W}\,Z_\mu\sum_f\bar f\gamma^\mu
\Big[T_3^f\big(P_L+\GR P_R\big)-Q_f s_W^2\Big]f
=\frac{g}{c_W}\,Z_\mu\sum_f\bar f\gamma^\mu
\big(g_V^{f,{\rm eff}}-g_A^{f,{\rm eff}}\gamma_5\big)f ,
\label{eq:LNCG}
\end{equation}
with
\begin{equation}
g_V^{f,{\rm eff}}=\tfrac12T_3^f\big(1+\GR\big)-Q_fs_W^2,\qquad
g_A^{f,{\rm eff}}=\tfrac12T_3^f\big(1-\GR\big).
\label{gvga}
\end{equation}
For $\GR=1$ the isospin part of the coupling becomes vector-like and parity
violation in the neutral current disappears. Eq.~\eqref{eq:LNCG} follows from writing the right-handed term in the
unbroken basis, $\mathscr{G}_R^Z\,\bar f\gamma^\mu P_R\big[g\,W^3_\mu T_3
+g'B_\mu(Y_L-Y_R)\big]f$, and rotating to the mass eigenstates
$W^3=c_WZ+s_WA$, $B=-s_WZ+c_WA$. Since $Y_L-Y_R=-T_3$, the right-handed
term couples only to the $Z$, $\propto(g/c_W)\,\mathscr{G}_R^Z T_3$, while
the photon coupling remains $eQ$ and electromagnetic current conservation
is unaffected.

\subsection{Weak charge}
Inserting Eq.~\eqref{gvga} into Eq.~\eqref{eq:C1C2} gives (see Appendix~\ref{appA}) the exact tree-level
couplings
\begin{equation}
C_{1q}=-T_3^q\big(1-\mathscr{G}^2\big)+2Q_q\,s_W^2\big(1-\mathscr{G}\big),\qquad
C_{2q}=-T_3^q\big(1-\mathscr{G}^2\big)+4T_3^q\,s_W^2\big(1-\mathscr{G}\big),
\label{eq:C1G}
\end{equation}
where $\mathscr{G}\equiv\GR(q^2)$ at the momentum transfer of the process. Note
that both the electron axial factor $(1-\mathscr{G})$ and the quark vector factor
$(1+\mathscr{G})$ must be kept: the pure-isospin pieces combine into
$(1-\mathscr{G})(1+\mathscr{G})=1-\mathscr{G}^2$ and cancel at first order.
Hence, to first order in $\GR$,
\begin{equation}
\boxed{\;s_W^2\;\longrightarrow\;s_W^2\big(1-\GR\big)\;}
\label{eq:rule}
\end{equation}
in every low-energy parity-violating coupling $C_{1q}$, $C_{2q}$ and in the
electron weak charge measured in M{\o}ller scattering; the same rule holds for the
Paschos--Wolfenstein combination of neutrino--nucleon scattering. For the nuclear
weak charge, Eq.~\eqref{eq:QW} becomes
\begin{equation}
Q_W(Z,N)=\big(1-\mathscr{G}^2\big)(Z-N)-4Z\,s_W^2\big(1-\mathscr{G}\big)
\simeq Z(1-4s_W^2)-N+4s_W^2\,Z\,\GR ,
\label{eq:QWG}
\end{equation}
i.e.\ the shift is proportional to the proton number only. For cesium, with
$\hat s^2(0)=0.23873$~\cite{PDG2024},
\begin{equation}
\Delta Q_W({\rm Cs})=52.5\,\GR(q^2\simeq0).
\label{eq:dQWCs}
\end{equation}
Comparing with Eq.~\eqref{eq:dQW}, the right-handed coupling corresponds to
$f_{Vq}/\Lambda^2=-\sqrt2\,G_F\,Q_q\,s_W^2\,\GR$, i.e.\
$f_{Vu}/\Lambda^2=-\tfrac23\sqrt2G_Fs_W^2\GR$ and
$f_{Vd}/\Lambda^2=+\tfrac13\sqrt2G_Fs_W^2\GR$: the new interaction couples to
the quark electric charge.

Using the value of Ref.~\cite{Dzuba},
\begin{equation}
\GR(q^2\simeq0)=0.013\pm0.008,\qquad \GR<0.029\ \ (95\%\ {\rm CL}),
\label{eq:GAPV}
\end{equation}
while the value of Ref.~\cite{Porsev} gives $\GR=0.002\pm0.007$ and
$\GR<0.015$. The difference illustrates that the APV bound is currently limited
by atomic theory. In the scenario considered here $\GR\to0$ at low energies, so
that Eq.~\eqref{eq:GAPV} is a consistency check of the infrared limit rather than a
strong constraint on the model.

\section{Effective weak mixing angle and running of $\GR$}\label{sec4}

\subsection{Left--right asymmetry at the $Z$ pole}
In $e^+e^-\to Z\to f\bar f$ the left--right asymmetry,
$A^0_{LR}\equiv(\sigma_L-\sigma_R)/(\sigma_L+\sigma_R)$, where $\sigma_{L(R)}$ is
the cross section for left- (right-)handed polarized electrons, measures the
electron couplings~\cite{ErlerSu,SLD}
\begin{equation}
A^0_{LR}=\frac{2g_V^{e,{\rm eff}}g_A^{e,{\rm eff}}}{(g_V^{e,{\rm eff}})^2+(g_A^{e,{\rm eff}})^2}
=\frac{1-4\sin^2\theta_{\rm eff}}{1-4\sin^2\theta_{\rm eff}+8\sin^4\theta_{\rm eff}},
\label{alrq}
\end{equation}
where the effective leptonic mixing angle is defined by
\begin{equation}
\sin^2\theta_{\rm eff}\equiv\frac14\left(1-\frac{g_V^{e,{\rm eff}}}{g_A^{e,{\rm eff}}}\right).
\label{seff}
\end{equation}
The same quantity is extracted from forward--backward asymmetries at LEP, the
Tevatron and the LHC~\cite{LEPSLD,Tevatron,CMS13,LHCb13}. Substituting Eq.~\eqref{gvga} (Appendix~\ref{appA4}) for
the electron gives the exact result
\begin{equation}
\sin^2\theta_{\rm eff}=\frac{s_W^2-\GR/2}{1-\GR}
\simeq s_W^2-\frac{\GR}{2}\big(1-2s_W^2\big).
\label{teff}
\end{equation}
The $Z$-pole observables are thus shifted by a different combination,
$-\tfrac12(1-2s_W^2)\GR$, than the low-energy ones,
$-s_W^2\GR$ [Eq.~\eqref{eq:rule}].

\subsection{Data-driven extraction of $\GR(q^2)$}

\begin{figure}[t]
\centering
\includegraphics[width=0.72\textwidth]{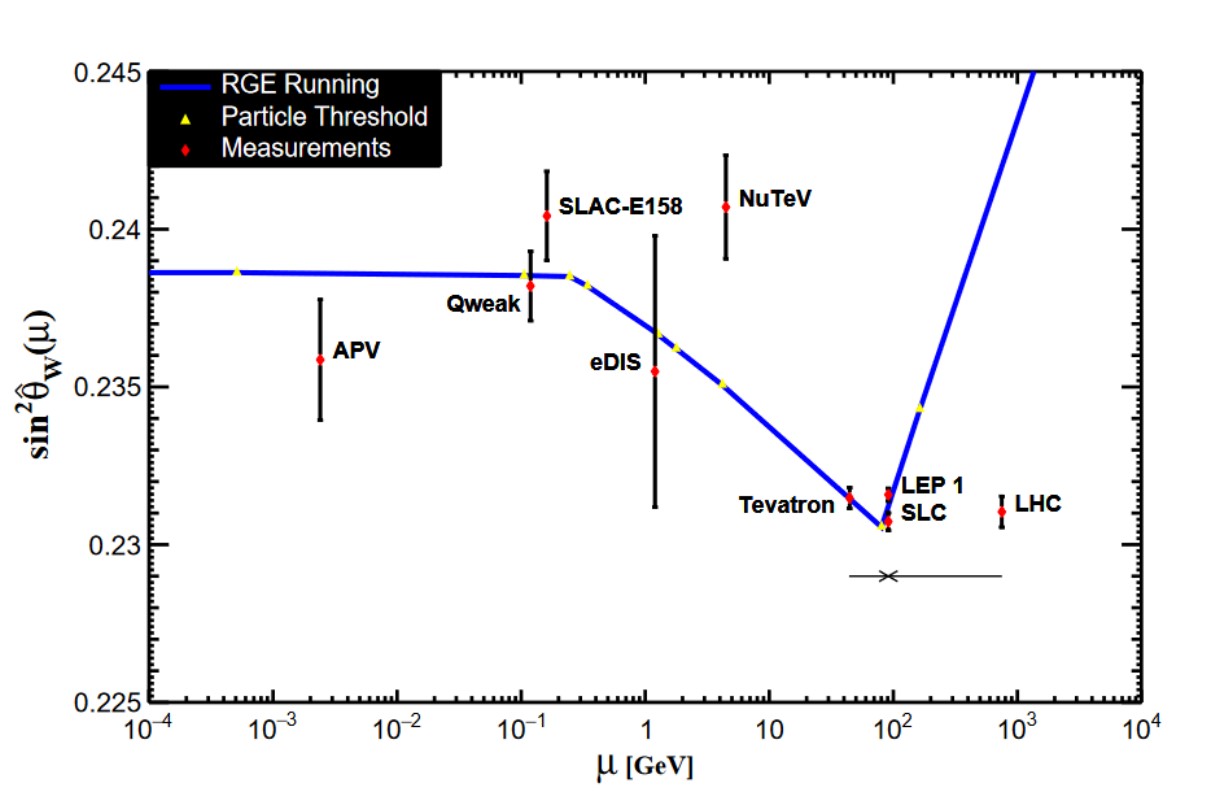}
\caption{Scale dependence of the weak mixing angle in the $\overline{\rm MS}$
scheme in the SM (solid line), together with its experimental determinations.
Figure reproduced from Ref.~\cite{PDG2024}. The values used in our analysis are
those of Table~\ref{tab:data}.}
\label{fig:running}
\end{figure}

In the SM the $\overline{\rm MS}$ weak mixing angle runs with the
scale~\cite{Erler2018,Jegerlehner}: $\hat s^2(\mu)\simeq0.2387$ for
$\mu\lesssim0.1$~GeV, decreasing to $\hat s^2(m_Z)=0.23129$~\cite{PDG2024} at the $Z$ pole and
increasing again above it (Fig.~\ref{fig:running}). Table~\ref{tab:data} lists
the determinations used in this work. Each provides a deviation
\begin{equation}
\Delta(q_i^2)\equiv\sin^2\theta^{\rm exp}(q_i^2)-\sin^2\theta^{\rm SM}(q_i^2),
\end{equation}
from which, by Eqs.~\eqref{eq:rule} and \eqref{teff},
\begin{equation}
\GR(q_i^2)=
\begin{cases}
-\Delta(q_i^2)/\hat s^2_{\rm SM}(q_i^2), & \text{low-energy PV observables},\\[4pt]
-2\Delta(m_Z^2)/\big(1-2\sin^2\theta^{\rm SM}_{\rm eff}\big), & Z\text{-pole }\sin^2\theta_{\rm eff}.
\end{cases}
\label{eq:extractG}
\end{equation}
For APV we use Eq.~\eqref{eq:dQWCs} directly. NuTeV~\cite{NuTeV} quotes the
on-shell mixing angle, $0.2277(16)$, to be compared with the SM on-shell value
$0.2227$; its $3\sigma$ deviation is widely attributed to nuclear and
parton-distribution effects, and we show it for completeness but exclude it from
the fit. For the SM prediction of $\sin^2\theta_{\rm eff}$ we use
$0.23161$~\cite{PDG2024}.

\begin{table}[t]
\centering
\small
\begin{tabular}{lccccc}
\toprule
Experiment & $\mu$ [GeV] & quantity & measured value & Ref. & $\GR$ \\
\midrule
APV (Cs) & $\sim0.0024$ & $Q_W$ & $-72.58\pm0.43$ & \cite{Dzuba} & $\phantom{-}0.013\pm0.008$\\
Qweak & 0.157 & $\hat s^2$ & $0.2382\pm0.0011$ & \cite{Qweak} & $\phantom{-}0.002\pm0.005$\\
SLAC E158 & 0.161 & $\hat s^2$ & $0.2403\pm0.0013$ & \cite{E158} & $-0.007\pm0.005$\\
PVDIS ($e$--D) & 1--1.4 & $\hat s^2$ & $0.2299\pm0.0043$ & \cite{PVDIS} & $\phantom{-}0.031\pm0.018$\\
NuTeV$^\ast$ & $\sim4$ & $s^2_{\rm on\text{-}shell}$ & $0.2277\pm0.0016$ & \cite{NuTeV} & $-0.021\pm0.007$\\
\midrule
LEP+SLD & $m_Z$ & $\sin^2\theta_{\rm eff}$ & $0.23153\pm0.00016$ & \cite{LEPSLD} & $(0.3\pm0.6)\times10^{-3}$\\
Tevatron & $m_Z$ & $\sin^2\theta_{\rm eff}$ & $0.23148\pm0.00033$ & \cite{Tevatron} & $(0.5\pm1.2)\times10^{-3}$\\
 {CMS (13 TeV)} & $m_Z$ & $\sin^2\theta_{\rm eff}$ &  {$0.23157\pm0.00031$} & \cite{CMS13} &  {$(0.2\pm1.2)\times10^{-3}$}\\
 {LHCb (13 TeV)} & $m_Z$ & $\sin^2\theta_{\rm eff}$ &  {$0.23152\pm0.00049$} & \cite{LHCb13} &  {$(0.3\pm1.8)\times10^{-3}$}\\
\bottomrule
\end{tabular}
\caption{Determinations of the weak mixing angle used in this work and the
corresponding values of $\GR$ from Eq.~\eqref{eq:extractG}. $\hat s^2$ denotes
the $\overline{\rm MS}$ value at the scale of the measurement.
 {For LHCb the statistical, systematic and theoretical uncertainties are
added in quadrature. The LHC measurements are treated as uncorrelated with each
other and with the Tevatron; correlated PDF uncertainties would slightly weaken
the combination.}}
\label{tab:data}
\end{table}

\begin{figure}[t]
\centering
\includegraphics[width=0.72\textwidth]{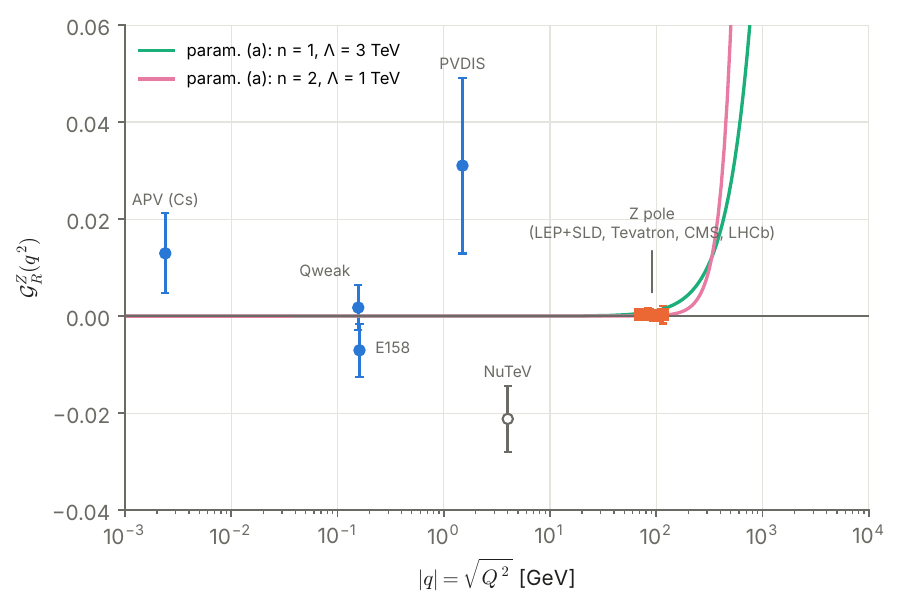}
\caption{Values of $\GR$ extracted from the measurements of
Table~\ref{tab:data} with Eq.~\eqref{eq:extractG}: low-energy observables
(circles), $Z$-pole $\sin^2\theta_{\rm eff}$ (squares) and NuTeV (open circle,
not used in the fits). The curves show the parametrization of
Eq.~\eqref{eq:Gparam} for two parameter choices close to the 95\% CL boundary of
Fig.~\ref{fig:excl}.}
\label{fig:GR}
\end{figure}

The extracted values are shown in Fig.~\ref{fig:GR}. All of them are consistent
with $\GR=0$ within about $1.5\sigma$ (NuTeV excepted). The $Z$-pole
determinations dominate; their weighted average is
\begin{equation}
\GR(m_Z^2)=(3.0\pm4.7)\times10^{-4},\qquad
\GR(m_Z^2)<1.1\times10^{-3}\ \ (95\%\ {\rm CL}),
\label{eq:GZ}
\end{equation}
consistent with the bound obtained from the $Z$ mass and width in
Ref.~\cite{Xue2022}. The low-energy points, by contrast, constrain $\GR$ only at
the percent level, as expected from their larger relative uncertainties.

\subsection{Constraints on the running of $\GR$}

\begin{figure}[t]
\centering
\includegraphics[width=\textwidth]{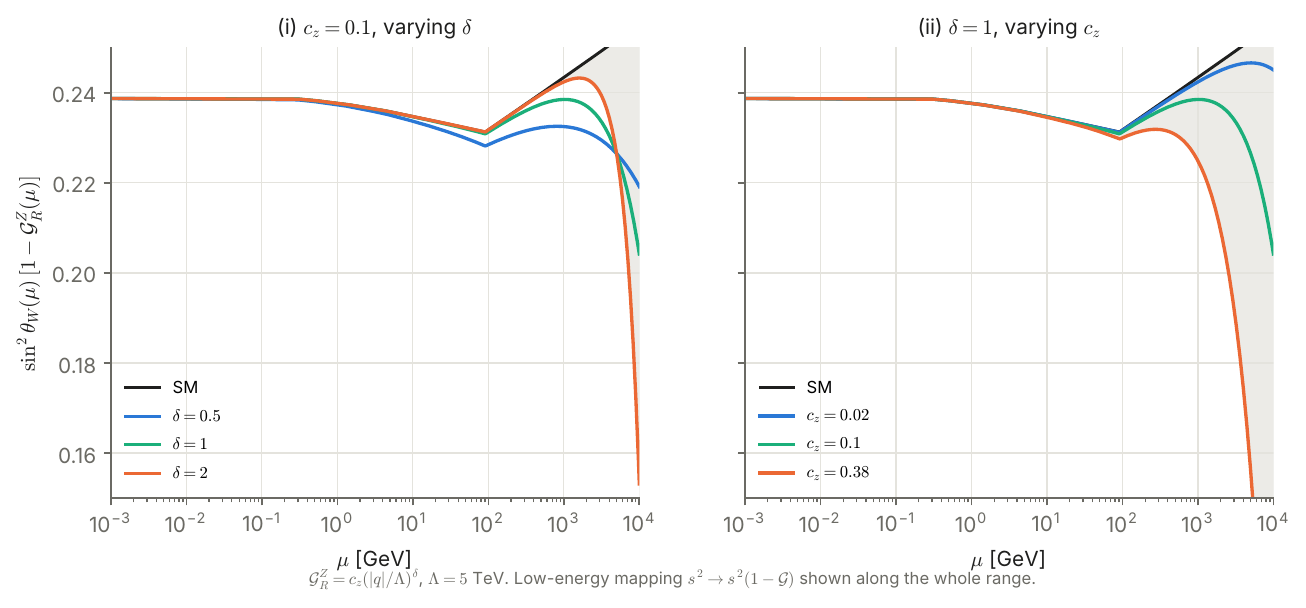}
\caption{Apparent low-energy weak mixing angle $\hat s^2(\mu)[1-\GR(\mu)]$ for
parametrization (b) of Eq.~\eqref{eq:Gparam} with $\Lambda=5$~TeV. Left:
$c_z=0.1$ and $\delta=0.5,1,2$. Right: $\delta=1$ and $c_z=0.02,0.1,0.38$. The
shaded area is spanned by the curves in each panel. The values $c_z\geq0.1$
with $\delta=1$ are excluded (Fig.~\ref{fig:excl}); they are shown for
illustration only.}
\label{fig:bands}
\end{figure}

To describe the transition between the parity-violating infrared regime and the
parity-restoring ultraviolet regime we consider two parametrizations,
\begin{equation}
\text{(a)}\quad \GR(q^2)=\frac{(q^2)^n}{(q^2)^n+\Lambda^{2n}},
\qquad\qquad
\text{(b)}\quad \GR(q^2)=c_z\left(\frac{|q|}{\Lambda}\right)^{\delta},
\label{eq:Gparam}
\end{equation}
where (a) satisfies $\GR\to0$ for $q^2\ll\Lambda^2$ and $\GR\to1$ for
$q^2\gg\Lambda^2$, and (b) is the low-energy form used in Ref.~\cite{Xue2022},
valid for $|q|\ll\Lambda$. Since all data are consistent with zero, they do not
determine the parameters but bound them from below. Applying Eq.~\eqref{eq:GZ}
at $q^2=m_Z^2$ alone (the high-mass Drell--Yan data are added in
Sec.~\ref{sec5}):
\begin{itemize}
\item for (a), $\Lambda>m_Z
\big[\frac{1-{\GR}^{\max}}{{\GR}^{\max}}\big]^{1/(2n)}$, i.e.\
$\Lambda>2.8$, $0.50$ and $0.28$~TeV for $n=1,2,3$;
\item for (b) with $\delta=1$, $c_z/\Lambda<1.2\times10^{-2}$~TeV$^{-1}$. For
$\Lambda=5$~TeV this requires $c_z<0.06$; the value $c_z=0.38$ of
Ref.~\cite{Xue2022} is excluded unless $\Lambda\gtrsim32$~TeV.
\end{itemize}
For (a) with $n\geq1$ the $Z$-pole bound weakens rapidly with $n$, because
$\GR(m_Z^2)\propto(m_Z/\Lambda)^{2n}$; measurements well above the $Z$ pole are
then needed. Fig.~\ref{fig:bands}
illustrates how the apparent low-energy mixing angle,
$\hat s^2(\mu)\,[1-\GR(\mu)]$, would deviate from the SM running for several
choices of the parameters of (b). Deviations become visible only above the $Z$
pole, where precision measurements of parity-violating neutral currents are not
yet available; Sec.~\ref{sec5} discusses how this region can be probed.

\section{TeV-scale constraints from high-mass Drell--Yan}\label{sec5}

\subsection{Forward--backward asymmetry with a right-handed coupling}
The most direct test of parity restoration at TeV scales is the forward--backward
asymmetry $A_{FB}$ of Drell--Yan lepton pairs at large invariant mass
$m_{\ell\ell}=\sqrt{q^2}$. For $q\bar q\to\gamma^*/Z\to\ell^-\ell^+$, with $\theta$
the angle between $\ell^-$ and the incoming quark in the dilepton rest frame,
\begin{equation}
\frac{d\hat\sigma_q}{d\cos\theta}\propto
\big(|\mathcal{A}^q_{LL}|^2+|\mathcal{A}^q_{RR}|^2\big)(1+\cos\theta)^2
+\big(|\mathcal{A}^q_{LR}|^2+|\mathcal{A}^q_{RL}|^2\big)(1-\cos\theta)^2 ,
\label{eq:dsdc}
\end{equation}
 {where the helicity amplitudes are}
\begin{equation}
\mathcal{A}^q_{ij}=Q_qQ_\ell+
\frac{g_i^q\,g_j^\ell}{s_W^2c_W^2}\,
\frac{q^2}{q^2-m_Z^2+i\,m_Z\Gamma_Z},\qquad i,j=L,R,
\label{eq:amps}
\end{equation}
 {and $g_L^f=T_3^f-Q_fs_W^2$ and $g_R^f=\GR\,T_3^f-Q_fs_W^2$ are the
coefficients of $P_L$ and $P_R$ in Eq.~\eqref{eq:LNCG}, i.e.\
$g_{L,R}^f=g_V^{f,{\rm eff}}\pm g_A^{f,{\rm eff}}$ with Eq.~\eqref{gvga}.}
The partonic asymmetry is
\begin{equation}
\hat A^q_{FB}=\frac34\,
\frac{|\mathcal{A}^q_{LL}|^2+|\mathcal{A}^q_{RR}|^2-|\mathcal{A}^q_{LR}|^2-|\mathcal{A}^q_{RL}|^2}
{\sum_{ij}|\mathcal{A}^q_{ij}|^2},
\label{eq:afbq}
\end{equation}
and the hadronic one is the average over quark flavours weighted by
$\mathcal{L}_{q\bar q}(q^2)\,\hat\sigma_q$. For $\GR\to1$ the isospin couplings
become vector-like, the $(1\pm\cos\theta)^2$ terms equalize and $A_{FB}$ falls
to a small value driven only by the $Q\,s_W^2$ terms. Above the $Z$ pole the
asymmetry is therefore very sensitive to $\GR$: we find
$\partial A_{FB}/\partial\GR\simeq-0.25$ at $m_{\ell\ell}=185$~GeV, rising
in magnitude to $-0.54$ above 1~TeV. Since the measured asymmetry is
$A_{FB}\simeq0.61$ with a precision of 2--5\% in each mass bin, one expects
sensitivity to $\GR$ at the level of a few percent, directly in the TeV range.

\subsection{Fit to the CMS data}
\begin{figure}[t]
\centering
\includegraphics[width=0.72\textwidth]{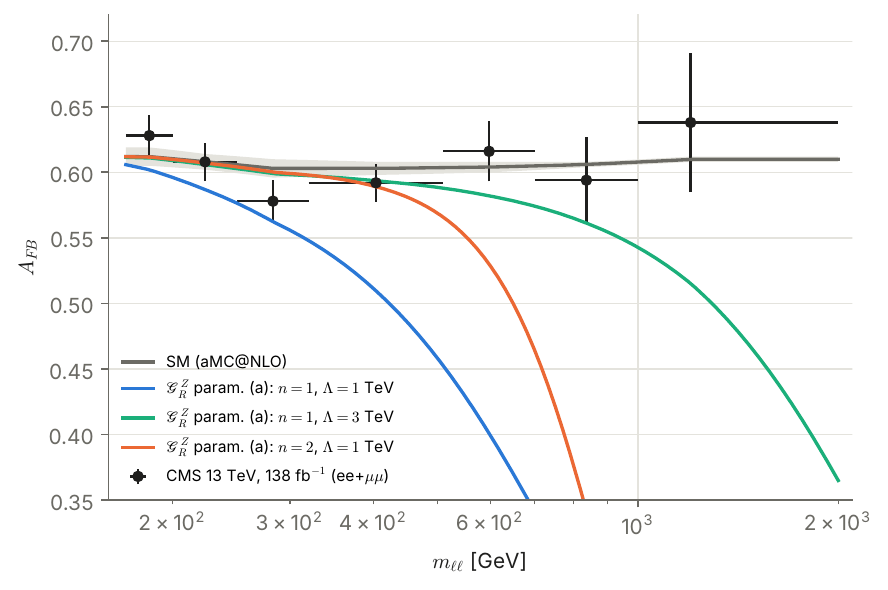}
\caption{CMS measurement of the Drell--Yan forward--backward asymmetry as a
function of $m_{\ell\ell}$~\cite{CMSAFB} compared with the SM (aMC@NLO, grey
band) and with the prediction of Eq.~\eqref{eq:afbth} for parametrization (a)
of Eq.~\eqref{eq:Gparam} with $(n,\Lambda)=(1,1~{\rm TeV})$, $(1,3~{\rm TeV})$
and $(2,1~{\rm TeV})$.}
\label{fig:afb}
\end{figure}

CMS has measured $A_{FB}$ in seven bins of $m_{\ell\ell}$ above 170~GeV, with
$138~{\rm fb}^{-1}$ at $\sqrt s=13$~TeV, corrected to the partonic
$\cos\theta^\ast$~\cite{CMSAFB}; the combined $ee+\mu\mu$ values and the
aMC@NLO predictions are listed in Table~\ref{tab:afb} and shown in Fig.~\ref{fig:afb}. We construct the
prediction as
\begin{equation}
A_{FB}^{\rm th}(m_{\ell\ell};\GR)=A_{FB}^{\rm aMC@NLO}(m_{\ell\ell})
+\Big[A_{FB}^{\rm LO}(m_{\ell\ell};\GR)-A_{FB}^{\rm LO}(m_{\ell\ell};0)\Big],
\label{eq:afbth}
\end{equation}
i.e.\ only the BSM shift is computed at leading order with
Eqs.~\eqref{eq:dsdc}--\eqref{eq:afbq}, so that higher-order QCD and
electroweak corrections and the photon-induced background are taken from the SM
prediction of CMS. We take $\GR$ universal for quarks and leptons, evaluate it at
the geometric centre of each bin (1.2~TeV for the last one), and use
$\sin^2\theta_{\rm eff}=0.2315$ and $\alpha(m_Z)=1/128$ in Eq.~\eqref{eq:amps}.
The ratio of down- to up-type parton luminosities, including sea quarks,
$r=\mathcal{L}_{d\bar d}/\mathcal{L}_{u\bar u}$, is set to 0.5 and varied
between 0.35 and 0.65. The $\chi^2$ includes statistical, systematic and
theoretical uncertainties. As the bin-to-bin correlation of the systematic
uncertainties is not given in Ref.~\cite{CMSAFB}, we consider the two extreme
cases of uncorrelated and fully correlated systematics.

\begin{table}[t]
\centering\small
\begin{tabular}{cccc}
\toprule
$m_{\ell\ell}$ [GeV] & $A_{FB}$ (CMS, $ee+\mu\mu$) & aMC@NLO & $\GR$ (this work)\\
\midrule
170--200 & $0.628\pm0.009\pm0.011$ & $0.612\pm0.007$ & $-0.06\pm0.06$\\
200--250 & $0.608\pm0.009\pm0.010$ & $0.608\pm0.006$ & $\phantom{-}0.00\pm0.04$\\
250--320 & $0.578\pm0.011\pm0.009$ & $0.603\pm0.007$ & $\phantom{-}0.06\pm0.04$\\
320--510 & $0.592\pm0.011\pm0.008$ & $0.603\pm0.005$ & $\phantom{-}0.02\pm0.03$\\
510--700 & $0.616\pm0.021\pm0.008$ & $0.604\pm0.004$ & $-0.02\pm0.04$\\
700--1000 & $0.594\pm0.032\pm0.008$ & $0.606\pm0.002$ & $\phantom{-}0.02\pm0.06$\\
$>1000$ & $0.638\pm0.052\pm0.011$ & $0.610\pm0.002$ & $-0.05\pm0.10$\\
\bottomrule
\end{tabular}
\caption{CMS measurement of $A_{FB}$ (stat.\ and syst.\ uncertainties) and the
aMC@NLO prediction~\cite{CMSAFB}, and the value of $\GR$ extracted in each mass
bin from Eq.~\eqref{eq:afbth}, linearized around $\GR=0$. The variation of $r$
changes the extracted values by less than $0.005$.}
\label{tab:afb}
\end{table}

\begin{figure}[t]
\centering
\includegraphics[width=0.72\textwidth]{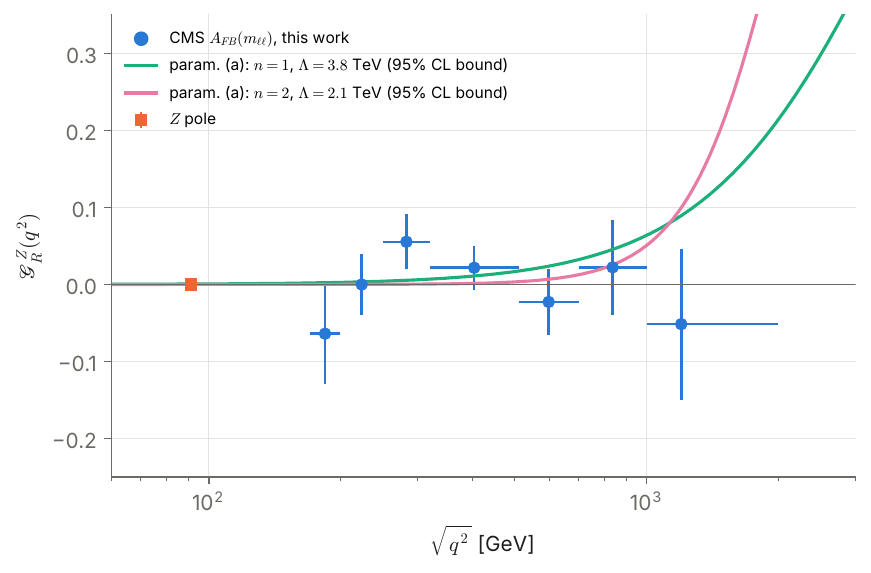}
\caption{$\GR(q^2)$ extracted from the CMS $A_{FB}$ in each mass bin
(circles; horizontal bars show the bin width) and from the $Z$-pole data
[Eq.~\eqref{eq:GZ}, square], together with parametrization (a) at its 95\% CL
lower bound on $\Lambda$ for $n=1$ and $n=2$.}
\label{fig:GRhigh}
\end{figure}

\subsection{Results}
The SM gives $\chi^2=4.8$ (7.2 with correlated systematics) for 7 bins, and the
bin-by-bin values of $\GR$ in Table~\ref{tab:afb} and Fig.~\ref{fig:GRhigh} are
all compatible with zero. For a constant right-handed coupling in the range
$0.17<\sqrt{q^2}<2$~TeV we find
\begin{equation}
\GR=0.011\pm0.015,\qquad \GR\big(0.2\text{--}2~{\rm TeV}\big)<0.04\ \ (95\%\ {\rm CL}),
\label{eq:Gconst}
\end{equation}
which weakens to $\GR<0.06$ for fully correlated systematics; the variation of
$r$ changes the limit by $\pm0.004$. To our knowledge this is the first direct
constraint on this coupling at TeV momentum transfer. It shows that
parity violation in the neutral current is maximal, to within 4\%, up to
$\sqrt{q^2}\approx2$~TeV.

\begin{figure}[t]
\centering
\includegraphics[width=\textwidth]{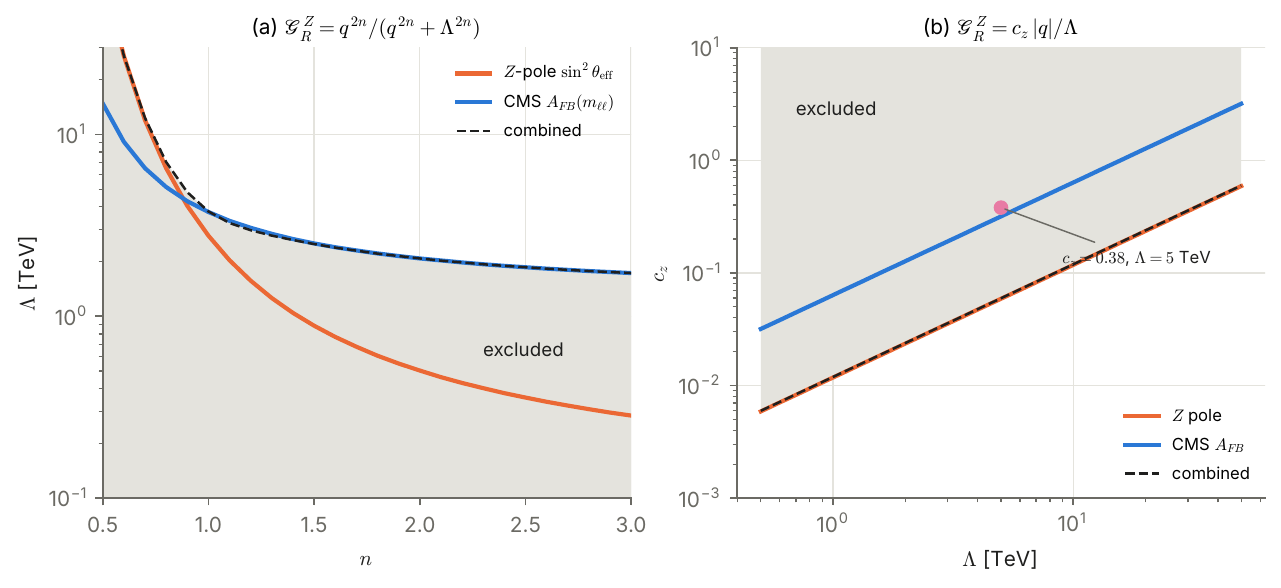}
\caption{Regions excluded at 95\% CL (shaded) by the $Z$-pole determinations of
$\sin^2\theta_{\rm eff}$, by the CMS high-mass $A_{FB}$, and by their combination.
Left: parametrization (a) of Eq.~\eqref{eq:Gparam} in the $(n,\Lambda)$ plane.
Right: parametrization (b) with $\delta=1$ in the $(\Lambda,c_z)$ plane; the dot
marks $c_z=0.38$, $\Lambda=5$~TeV.}
\label{fig:excl}
\end{figure}

For the running parametrizations of Eq.~\eqref{eq:Gparam} we combine the
$A_{FB}$ $\chi^2$ with Eq.~\eqref{eq:GZ} and set one-sided 95\% CL limits
($\Delta\chi^2=2.71$). The results are collected in Table~\ref{tab:lambda} and
Fig.~\ref{fig:excl}. For (a) the Drell--Yan data dominate for all $n\geq1$ and
improve the $Z$-pole bound by a factor of four for $n=2$ and six for $n=3$:
\begin{equation}
\Lambda>3.8~{\rm TeV}\ (n=1),\qquad \Lambda>2.1~{\rm TeV}\ (n=2),\qquad
\Lambda>1.7~{\rm TeV}\ (n=3).
\label{eq:Lambdabounds}
\end{equation}
These bounds change by less than 5\% when $r$ or the systematic correlations are
varied. For (b) with $\delta\leq1$ the $Z$ pole dominates, whereas for $\delta=2$
the Drell--Yan data are  {more} constraining. The value $c_z=0.38$ at
$\Lambda=5$~TeV with $\delta=1$, used in Ref.~\cite{Xue2022}, is excluded by the
Drell--Yan data alone ($c_z<0.32$) as well as by the $Z$ pole ($c_z<0.06$).

\begin{table}[t]
\centering\small
\begin{tabular}{lccc}
\toprule
 & $Z$ pole only & CMS $A_{FB}$ only & combined\\
\midrule
(a), $n=1$: $\Lambda>$ & 2.8 TeV & 3.7 TeV & 3.8 TeV\\
(a), $n=2$: $\Lambda>$ & 0.50 TeV & 2.1 TeV & 2.1 TeV\\
(a), $n=3$: $\Lambda>$ & 0.28 TeV & 1.7 TeV & 1.7 TeV\\
\midrule
(b), $\Lambda=5$ TeV, $\delta=0.5$: $c_z<$ & 0.008 & 0.11 & 0.008\\
(b), $\Lambda=5$ TeV, $\delta=1$: $c_z<$ & 0.059 & 0.32 & 0.060\\
(b), $\Lambda=5$ TeV, $\delta=2$: $c_z<$ & 3.2 & 1.6 & 1.6\\
\bottomrule
\end{tabular}
\caption{One-sided 95\% CL limits on the parameters of Eq.~\eqref{eq:Gparam}.}
\label{tab:lambda}
\end{table}

\subsection{Limitations}
Our analysis is at tree level in the new coupling and assumes a universal $\GR$
for all fermions. A non-zero $\GR(m_Z^2)$ would also modify the $Z$ partial
widths and hence other inputs of the global electroweak fit, which we do not
refit. A possible $q^2$ dependence of $\GR$ within a mass bin is neglected. The
right-handed couplings could also produce resonant or contact-interaction effects
in the dilepton mass spectrum; the asymmetry is largely insensitive to the
overall normalization and thus to such effects, which deserve a separate study.

\section{Future probes}\label{sec6}

\paragraph{Future low-energy experiments.}
The MOLLER experiment at Jefferson Lab~\cite{MOLLER} and P2 at
Mainz~\cite{P2} aim at determinations of $\hat s^2$ at $Q\sim0.1$~GeV with an
uncertainty of about $3\times10^{-4}$. By Eq.~\eqref{eq:rule} this corresponds to
a sensitivity to $\GR$ of order $10^{-3}$ at low momentum transfer, comparable to
the present $Z$-pole bound. Improved atomic-structure calculations for cesium,
and APV measurements in other atoms, would sharpen Eq.~\eqref{eq:GAPV}
correspondingly.

\paragraph{LHC Run 3 and HL-LHC.}
The statistical uncertainty of $A_{FB}$ above 500~GeV dominates the present
bound. With $3~{\rm ab}^{-1}$ at the HL-LHC the statistical errors decrease by a
factor of about 4.5, and we expect the sensitivity to $\GR$ in the TeV range
to reach the 1\% level, provided systematic uncertainties are controlled at a
similar level. 
\section{Summary and remarks}\label{secsumm}

We have computed the effect of a right-handed $Z$ coupling $\GR$, predicted by a
four-fermion-interaction scenario with sterile neutrinos and TeV-scale parity
restoration, on neutral-current parity-violating observables. Our main results
are:
\begin{enumerate}
\item At first order, all low-energy parity-violating couplings follow from the SM
by $s_W^2\to s_W^2(1-\GR)$; the nuclear weak charge is shifted by
$\Delta Q_W=4s_W^2Z\,\GR$ [Eq.~\eqref{eq:QWG}], i.e.\ $\Delta Q_W({\rm Cs})=52.5\,\GR$.
\item The $Z$-pole effective angle is shifted by $-\tfrac12(1-2s_W^2)\GR$
[Eq.~\eqref{teff}], so that low- and high-energy observables must be interpreted
with different relations [Eq.~\eqref{eq:extractG}].
\item APV in cesium gives $\GR(0)=0.013\pm0.008$ or $0.002\pm0.007$, depending
on the atomic-theory input; the $Z$-pole data give
$\GR(m_Z^2)<1.1\times10^{-3}$ at 95\% CL.
\item The CMS high-mass Drell--Yan forward--backward asymmetry constrains the
coupling directly at TeV momentum transfer: $\GR<0.04$ (95\% CL) for
$0.2<\sqrt{q^2}<2$~TeV.
\item For a running $\GR=(q^2)^n/[(q^2)^n+\Lambda^{2n}]$ the scale of parity
restoration must satisfy $\Lambda>3.8$, $2.1$ and $1.7$~TeV for $n=1,2,3$;
the linear form with $c_z=0.38$, $\Lambda=5$~TeV is excluded.
\end{enumerate}
Higher-luminosity Drell--Yan data at the LHC and the forthcoming MOLLER and P2
experiments will test the scenario at TeV and at low momentum transfer,
respectively.

 {In the scenario considered here $\GR(q^2)$ increases toward unity as the
momentum transfer grows, so that the left- and right-handed isospin couplings
become equal and parity symmetry is restored in the neutral current. The present
data are compatible with this picture provided the transition sets in above
$\Lambda\simeq2$--$4$~TeV (Table~\ref{tab:lambda}). They are not yet precise
enough to observe the growth itself; detecting it is the main target of future
high-luminosity Drell--Yan measurements.}

\appendix
\section{Derivation of the effective couplings with $\GR$}\label{appA}

In this appendix we give the intermediate steps leading to
Eqs.~\eqref{eq:C1C2}, \eqref{eq:C1G}, \eqref{eq:QWG} and \eqref{teff}. We write
$\mathscr{G}\equiv\GR(q^2)$ and take it universal for quarks and leptons.

\subsection{Four-fermion couplings}\label{appA1}
Starting from $\mathcal{L}_{\rm NC}=(g/c_W)Z_\mu J_Z^\mu$ with
$J_Z^\mu=\sum_f\bar f\gamma^\mu(g_V^f-g_A^f\gamma_5)f$, the equation of motion of
the $Z$ at $|q^2|\ll m_Z^2$ gives $Z_\mu=-(g/c_W)J_{Z\mu}/m_Z^2$, hence
\begin{equation}
\mathcal{L}_{\rm eff}=-\frac{g^2}{2c_W^2m_Z^2}\,J_{Z\mu}J_Z^\mu
=-2\sqrt2\,G_F\,J_{Z\mu}J_Z^\mu ,
\label{eq:A1}
\end{equation}
where $g^2/(8c_W^2m_Z^2)=G_F/\sqrt2$ was used. The electron--quark part of
$J_{Z\mu}J_Z^\mu$ appears twice. Its axial-electron $\times$ vector-quark piece is
\begin{equation}
-2\sqrt2\,G_F\cdot2\,(-g_A^e)\,g_V^q\;
\bar e\gamma^\mu\gamma_5e\;\bar q\gamma_\mu q
=4\sqrt2\,G_F\,g_A^eg_V^q\;\bar e\gamma^\mu\gamma_5e\;\bar q\gamma_\mu q ,
\end{equation}
and similarly for the vector-electron $\times$ axial-quark piece. Comparing with
Eq.~\eqref{eq:Leff} gives $C_{1q}=8g_A^eg_V^q$ and $C_{2q}=8g_V^eg_A^q$,
Eq.~\eqref{eq:C1C2}.

\subsection{Electron couplings}\label{appA2}
With $T_3^e=-\tfrac12$ and $Q_e=-1$, Eq.~\eqref{gvga} gives
\begin{equation}
g_A^e=-\tfrac14(1-\mathscr{G}),\qquad
g_V^e=-\tfrac14(1+\mathscr{G})+s_W^2 .
\label{eq:A3}
\end{equation}

\subsection{The coefficients $C_{1q}$ and $C_{2q}$}\label{appA3}
Inserting Eq.~\eqref{eq:A3} and the quark couplings of Eq.~\eqref{gvga} into
Eq.~\eqref{eq:C1C2},
\begin{align}
C_{1q}&=8\Big[-\tfrac14(1-\mathscr{G})\Big]\Big[\tfrac12T_3^q(1+\mathscr{G})-Q_qs_W^2\Big]
=-T_3^q(1-\mathscr{G})(1+\mathscr{G})+2Q_q\,s_W^2(1-\mathscr{G}),\\
C_{2q}&=8\Big[-\tfrac14(1+\mathscr{G})+s_W^2\Big]\Big[\tfrac12T_3^q(1-\mathscr{G})\Big]
=-T_3^q(1+\mathscr{G})(1-\mathscr{G})+4T_3^q\,s_W^2(1-\mathscr{G}),
\end{align}
which is Eq.~\eqref{eq:C1G}. For $\mathscr{G}=0$ these reduce to the SM values,
as Table~\ref{tab:A1} shows.

\begin{table}[h]
\centering\small
\begin{tabular}{cccc}
\toprule
$q$ & $(T_3,\,Q)$ & $C_{1q}$ & $C_{2q}$\\
\midrule
$u$ & $(+\tfrac12,\,+\tfrac23)$ & $-\tfrac12+\tfrac43s_W^2$ & $-\tfrac12+2s_W^2$\\
$d$ & $(-\tfrac12,\,-\tfrac13)$ & $+\tfrac12-\tfrac23s_W^2$ & $+\tfrac12-2s_W^2$\\
\bottomrule
\end{tabular}
\caption{SM limit ($\mathscr{G}=0$) of Eq.~\eqref{eq:C1G}.}
\label{tab:A1}
\end{table}

The important point is that the isospin terms contain the product
$(1-\mathscr{G})(1+\mathscr{G})$. One factor comes from the axial coupling of
one fermion and the other from the vector coupling of the other, so the linear
term cancels. If only the quark factor $(1+\mathscr{G})$ is kept, one finds
instead $C_{1q}\simeq C_{1q}^{\rm SM}-T_3^q\mathscr{G}$, which is incorrect. To first
order,
\begin{equation}
C_{1q}\simeq-T_3^q+2Q_q\big[s_W^2(1-\mathscr{G})\big],\qquad
C_{2q}\simeq-T_3^q+4T_3^q\big[s_W^2(1-\mathscr{G})\big],
\end{equation}
which is the SM form with $s_W^2\to s_W^2(1-\mathscr{G})$, Eq.~\eqref{eq:rule}.
The same argument applies to any low-energy parity-violating amplitude, which is
always proportional to a product $g_A^ag_V^b$:
\begin{equation}
g_A^ag_V^b=\frac{T_3^aT_3^b}{4}(1-\mathscr{G}^2)-\frac{T_3^aQ_b}{2}\,s_W^2(1-\mathscr{G}).
\end{equation}

\subsection{Nuclear weak charge}\label{appA5}
From Eq.~\eqref{eq:QW},
\begin{align}
Q_W&=-2\Big\{(2Z+N)\Big[-\tfrac12(1-\mathscr{G}^2)+\tfrac43s_W^2(1-\mathscr{G})\Big]
+(Z+2N)\Big[\tfrac12(1-\mathscr{G}^2)-\tfrac23s_W^2(1-\mathscr{G})\Big]\Big\}\nonumber\\
&=-2\Big\{\tfrac12(1-\mathscr{G}^2)\big[(Z+2N)-(2Z+N)\big]
+\tfrac23s_W^2(1-\mathscr{G})\big[2(2Z+N)-(Z+2N)\big]\Big\}\nonumber\\
&=-2\Big\{\tfrac12(1-\mathscr{G}^2)(N-Z)+2Z\,s_W^2(1-\mathscr{G})\Big\}\nonumber\\
&=(1-\mathscr{G}^2)(Z-N)-4Z\,s_W^2(1-\mathscr{G}),
\end{align}
where we used $(Z+2N)-(2Z+N)=N-Z$ and $2(2Z+N)-(Z+2N)=3Z$. Expanding to first
order gives Eq.~\eqref{eq:QWG}, $\Delta Q_W=4s_W^2Z\,\mathscr{G}$. The shift
depends only on the proton number, because the $s_W^2$ term couples to the
electric charge of the quarks. Equivalently, the weak
charge of the neutron, $Q_W^n=-(1-\mathscr{G}^2)$, is unchanged at first order.

\subsection{Effective mixing angle at the $Z$ pole}\label{appA4}
Inserting Eq.~\eqref{eq:A3} into Eq.~\eqref{seff},
\begin{equation}
\sin^2\theta_{\rm eff}=\frac14\left[1-\frac{-\tfrac14(1+\mathscr{G})+s_W^2}{-\tfrac14(1-\mathscr{G})}\right]
=\frac14\left[1-\frac{(1+\mathscr{G})-4s_W^2}{1-\mathscr{G}}\right]
=\frac{s_W^2-\mathscr{G}/2}{1-\mathscr{G}} .
\end{equation}
Expanding, $(s_W^2-\mathscr{G}/2)(1+\mathscr{G})+\mathcal{O}(\mathscr{G}^2)
=s_W^2-\tfrac12(1-2s_W^2)\mathscr{G}$, which is Eq.~\eqref{teff}. Here
$g_V^e/g_A^e$ is a ratio of couplings of the same fermion, so the factors
$(1\pm\mathscr{G})$ do not cancel. This is why $Z$-pole and low-energy
observables respond differently to $\mathscr{G}$.



\begin{thebibliography}{99}\small
\bibitem{LeeYang} T.~D.~Lee and C.~N.~Yang, Phys.\ Rev.\ \textbf{104} (1956) 254. \doilink{10.1103/PhysRev.104.254}
\bibitem{Wu} C.~S.~Wu \emph{et al.}, Phys.\ Rev.\ \textbf{105} (1957) 1413. \doilink{10.1103/PhysRev.105.1413}
\bibitem{Haidt} D.~Haidt, Eur.\ Phys.\ J.\ C \textbf{34} (2004) 25. \inspirelink{Eur.Phys.J.C,34,25}
\bibitem{Prescott} C.~Y.~Prescott \emph{et al.}, Phys.\ Lett.\ B \textbf{77} (1978) 347; Phys.\ Lett.\ B \textbf{84} (1979) 524. \inspirelink{Phys.Lett.B,77,347}; \inspirelink{Phys.Lett.B,84,524}
\bibitem{Wood} C.~S.~Wood \emph{et al.}, Science \textbf{275} (1997) 1759. \doilink{10.1126/science.275.5307.1759}
\bibitem{ErlerSu} J.~Erler and S.~Su, Prog.\ Part.\ Nucl.\ Phys.\ \textbf{71} (2013) 119. \arxivlink{1303.5522}
\bibitem{Bouchiat81} C.~Bouchiat, \emph{Atomic Physics 7} (1981).
\bibitem{Bouchiat97} M.-A.~Bouchiat and C.~Bouchiat, Rep.\ Prog.\ Phys.\ \textbf{60} (1997) 1351. \inspirelink{Rept.Prog.Phys.,60,1351}
\bibitem{RamseyMusolf} M.~J.~Ramsey-Musolf, Phys.\ Rev.\ C \textbf{60} (1999) 015501. \doilink{10.1103/PhysRevC.60.015501}
\bibitem{Rosner} J.~L.~Rosner, Phys.\ Rev.\ D \textbf{61} (1999) 016006. \doilink{10.1103/PhysRevD.61.016006}
\bibitem{Xue2016} S.-S.~Xue, JHEP \textbf{11} (2016) 072. \doilink{10.1007/JHEP11(2016)072}
\bibitem{Xue2017} S.-S.~Xue, JHEP \textbf{05} (2017) 146. \doilink{10.1007/JHEP05(2017)146}
\bibitem{Xue2022} S.-S.~Xue, Nucl.\ Phys.\ B \textbf{985} (2022) 115992. \doilink{10.1016/j.nuclphysb.2022.115992}
\bibitem{BouchiatFayet} C.~Bouchiat and P.~Fayet, Phys.\ Lett.\ B \textbf{608} (2005) 87. \arxivlink{hep-ph/0410260}
\bibitem{Arcadi} G.~Arcadi, M.~Lindner, J.~Martins and F.~S.~Queiroz, Nucl.\ Phys.\ B \textbf{959} (2020) 115158. \doilink{10.1016/j.nuclphysb.2020.115158}
\bibitem{PDG2024} S.~Navas \emph{et al.} (Particle Data Group), Phys.\ Rev.\ D \textbf{110} (2024) 030001. \doilink{10.1103/PhysRevD.110.030001}
\bibitem{Dzuba} V.~A.~Dzuba, J.~C.~Berengut, V.~V.~Flambaum and B.~Roberts, Phys.\ Rev.\ Lett.\ \textbf{109} (2012) 203003. \doilink{10.1103/PhysRevLett.109.203003}
\bibitem{Porsev} S.~G.~Porsev, K.~Beloy and A.~Derevianko, Phys.\ Rev.\ Lett.\ \textbf{102} (2009) 181601. \doilink{10.1103/PhysRevLett.102.181601}
\bibitem{Xue2013} S.-S.~Xue, Phys.\ Lett.\ B \textbf{727} (2013) 308. \inspirelink{Phys.Lett.B,727,308}
\bibitem{Xue2016b} S.-S.~Xue, Phys.\ Rev.\ D \textbf{93} (2016) 073001. \doilink{10.1103/PhysRevD.93.073001}
\bibitem{CDF} T.~Aaltonen \emph{et al.} (CDF Collaboration), Science \textbf{376} (2022) 170. \doilink{10.1126/science.abk1781}
\bibitem{ATLASW} G.~Aad \emph{et al.} (ATLAS Collaboration), Eur.\ Phys.\ J.\ C \textbf{84} (2024) 1309, \arxivlink{2403.15085}.
\bibitem{CMSW} CMS Collaboration, \arxivlink{2412.13872}.
\bibitem{CMSVBF} A.~Tumasyan \emph{et al.} (CMS Collaboration), Phys.\ Rev.\ Lett.\ \textbf{131} (2023) 011803. \doilink{10.1103/PhysRevLett.131.011803}
\bibitem{CMSHNL1} A.~M.~Sirunyan \emph{et al.} (CMS Collaboration), Phys.\ Rev.\ Lett.\ \textbf{120} (2018) 221801. \doilink{10.1103/PhysRevLett.120.221801}
\bibitem{CMSHNL2} A.~M.~Sirunyan \emph{et al.} (CMS Collaboration), JHEP \textbf{01} (2019) 122. \doilink{10.1007/JHEP01(2019)122}
\bibitem{Pacioselli} L.~Pacioselli, O.~Panella, M.~Presilla and S.-S.~Xue, JHEP \textbf{11} (2023) 054. \doilink{10.1007/JHEP11(2023)054}
\bibitem{Mahmoudi} S.~Mahmoudi \emph{et al.}, Eur.\ Phys.\ J.\ C \textbf{84} (2024) 619. \doilink{10.1140/epjc/s10052-024-13004-0}
\bibitem{Motie} I.~Motie \emph{et al.}, PTEP \textbf{2025} (2025) 113B03. \arxivlink{2409.07184}
\bibitem{Erler2018} J.~Erler and R.~Ferro-Hern\'andez, JHEP \textbf{03} (2018) 196. \doilink{10.1007/JHEP03(2018)196}
\bibitem{SLD} K.~Abe \emph{et al.} (SLD Collaboration), Phys.\ Rev.\ Lett.\ \textbf{84} (2000) 5945. \doilink{10.1103/PhysRevLett.84.5945}
\bibitem{LEPSLD} ALEPH, DELPHI, L3, OPAL, SLD Collaborations and LEP/SLD Electroweak Working Groups, Phys.\ Rept.\ \textbf{427} (2006) 257. \doilink{10.1016/j.physrep.2005.12.006}
\bibitem{Tevatron} T.~Aaltonen \emph{et al.} (CDF and D0 Collaborations), Phys.\ Rev.\ D \textbf{97} (2018) 112007. \doilink{10.1103/PhysRevD.97.112007}
\bibitem{CMS13} A.~Hayrapetyan \emph{et al.} (CMS Collaboration), \arxivlink{2408.07622} [hep-ex].
\bibitem{LHCb13} R.~Aaij \emph{et al.} (LHCb Collaboration), JHEP \textbf{12} (2024) 026. \doilink{10.1007/JHEP12(2024)026}
\bibitem{Jegerlehner} F.~Jegerlehner, Nuovo Cim.\ C \textbf{034S1} (2011) 31. \inspirelink{Nuovo+Cim.C,34S1,31}
\bibitem{NuTeV} G.~P.~Zeller \emph{et al.} (NuTeV Collaboration), Phys.\ Rev.\ Lett.\ \textbf{88} (2002) 091802. \doilink{10.1103/PhysRevLett.88.091802}
\bibitem{Qweak} D.~Androi\'c \emph{et al.} (Qweak Collaboration), Nature \textbf{557} (2018) 207. \doilink{10.1038/s41586-018-0096-0}
\bibitem{E158} P.~L.~Anthony \emph{et al.} (SLAC E158 Collaboration), Phys.\ Rev.\ Lett.\ \textbf{95} (2005) 081601. \doilink{10.1103/PhysRevLett.95.081601}
\bibitem{PVDIS} D.~Wang \emph{et al.} (PVDIS Collaboration), Nature \textbf{506} (2014) 67. \doilink{10.1038/nature12964}
\bibitem{CMSAFB} A.~Tumasyan \emph{et al.} (CMS Collaboration), JHEP \textbf{08} (2022) 063. \doilink{10.1007/JHEP08(2022)063}
\bibitem{MOLLER} J.~Benesch \emph{et al.} (MOLLER Collaboration), \arxivlink{1411.4088}.
\bibitem{P2} D.~Becker \emph{et al.}, Eur.\ Phys.\ J.\ A \textbf{54} (2018) 208. \inspirelink{Eur.Phys.J.A,54,208}
\end{thebibliography}
\end{document}